\documentclass[reprint,superscriptaddress,amsmath,amssymb,aps,pra,floatfix]{revtex4-1}
\usepackage[utf8]{inputenc}
\usepackage{amsmath}
\usepackage{amssymb}
\usepackage{physics}
\usepackage{enumerate}
\usepackage{graphicx}
\usepackage{float}
\usepackage{mathtools}
\usepackage{bbold}
\usepackage{comment}

\DeclareMathOperator\supp{supp}
\DeclarePairedDelimiter\set\{\}
\newcommand{\ba}{\begin{eqnarray}}
\newcommand{\ea}{\end{eqnarray}}
\newcommand{\ban}{\begin{eqnarray*}}
\newcommand{\ean}{\end{eqnarray*}}

\begin{document}

\title{Devices in Superposition}
\author{Asaph Ho}
\affiliation{Centre for Quantum Technologies, National University of Singapore, 3 Science Drive 2, Singapore 117543, Singapore}

\author{Valerio Scarani}
\affiliation{Department of Physics, National University of Singapore, 2 Science Drive 3, Singapore 117542, Singapore}
\affiliation{Centre for Quantum Technologies, National University of Singapore, 3 Science Drive 2, Singapore 117543, Singapore}

\begin{abstract}
    Under the assumption that every material object can ultimately be described by quantum theory, we ask how a probe system evolves in a device prepared and kept in a superposition state of values of its classical parameter. We find that, under ideal conditions, the evolution of the system would be unitary, generated by an effective Hamiltonian. We describe also an incoherent use of the device that achieves the same effective evolution on an ensemble. The effective Hamiltonian thus generated may have qualitatively different features from that associated to a classical value of the parameter.
\end{abstract}

\maketitle

\section{Introduction}

The superposition principle is one of the pillars of quantum mechanics. Given any two pure states of a system, each described by a vector, any linear superposition describes a valid pure state, which is different but somehow related to the initial two. Mathematically, this simply means that every vector is a valid state. However, even if two states correspond each to an intuitive classical narrative, their superposition usually does not (the qubit being the exception: any state can be associated to a direction in space). Schr\"odinger's deliberately paradoxical example of a cat in the superposition of being alive and being dead is often invoked as the paradigm of this situation \cite{schrodinger1935present}. Making sense of such superposition states or explaining them away has occupied a large fraction of the discussions on the meaning of quantum theory. Meanwhile, the observation of superposition in systems that could be called macroscopic \cite{Leggett_2002} has witnessed remarkable progress in the last two decades and is the subject of books and review articles \cite{Harochebook,RevModPhys.85.1103,arndt2014testing,RevModPhys.90.025004}. We are at a stage when one can create and detect superposition states of large molecules \cite{shayeghi2019matterwave} and almost micrometer-size mechanical oscillators \cite{ockeloen2018stabilized,riedinger2018remote,Deliceaba3993}.

In this paper we ask how a probe system (S) would perceive a device (D) prepared in a superposition state of what we would consider its natural (i.e.~classical) parameter. As examples, one may think of a scatterer prepared in a superposition state of several locations \cite{Schomerus_2002,PhysRevLett.96.173601}, a beam-splitter in superposition of different values of transmittivity \cite{PhysRevLett.115.260403, Angelo_2011,giacomini2019quantum}, a Stern-Gerlach magnet in superposition of different directions of magnetic field gradient, etc. Crucially to our proposal, the device should not only be \textit{prepared} in such a superposition state, but must also be \textit{maintained} in that state during the interaction with the system. This will be guaranteed by constant monitoring of the device, through the Zeno effect \cite{doi:10.1063/1.523304}. We prove that, in the limit of strong Zeno effect, the system experiences a \textit{unitary} evolution, generated by a Hamiltonian that depends on the state in which the device is frozen. We present cases in which this Hamiltonian is qualitatively different from the one associated to classical states of the device. 

Because the scheme requires both the preparation of a superposition state of the device and the Zeno freezing in that same state, an experimental realisation is probably a long shot away from existing technologies (although we do not want to underestimate the experimentalists' ingenuity). Nonetheless, contrary to recent ideas on superposition of time-ordering of evolutions \cite{chiribella2013quantum,brukner2014quantum}, our work fits within normal quantum theory, provided the superposition principle holds at any scale.

\begin{figure}[t]
    \centering
    \includegraphics[width=\columnwidth]{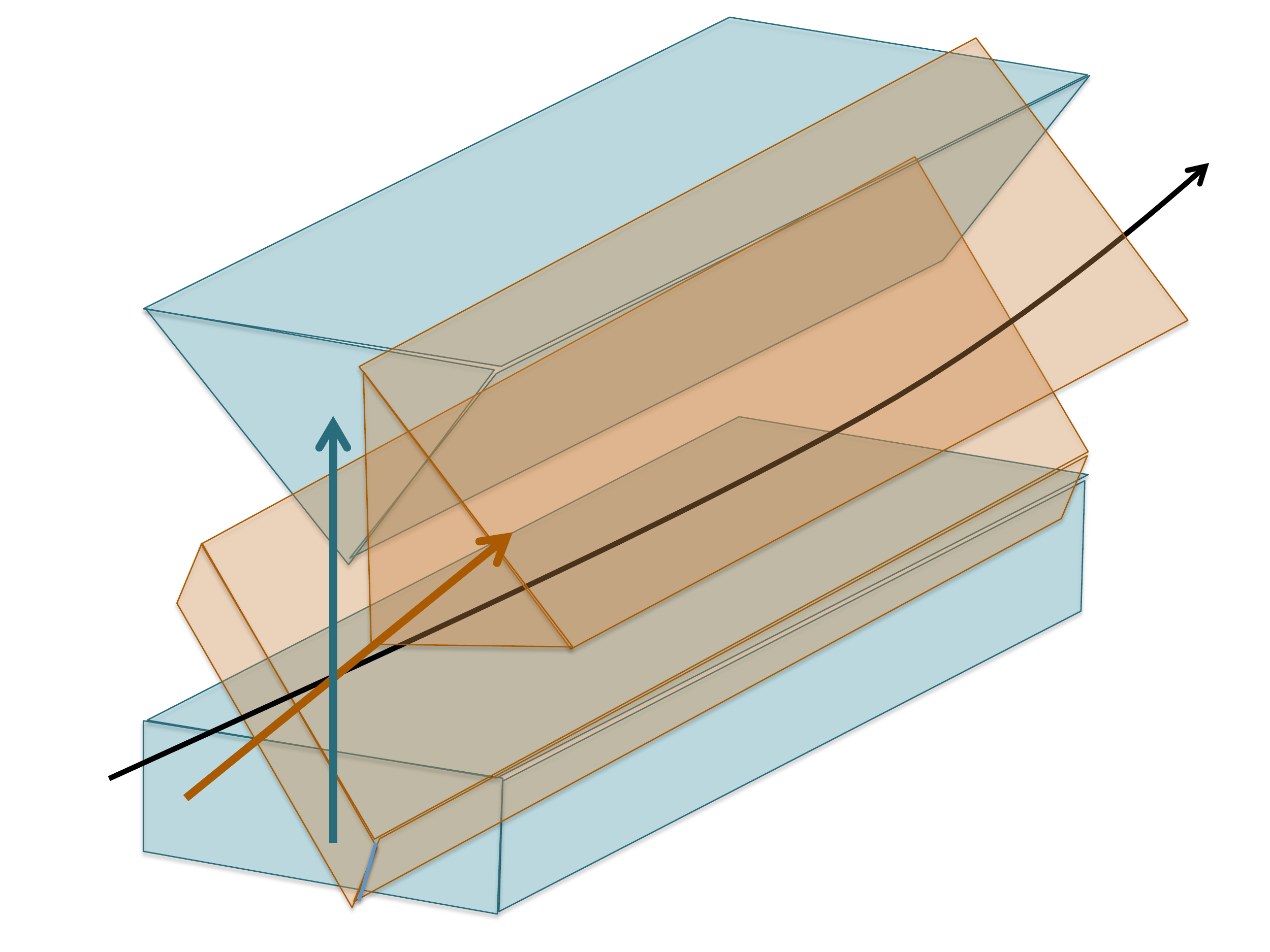}
    \caption{(Color online) Depiction of the question answered in this paper: how would a probe system (black trajectory) perceive a device (here, a Stern-Gerlach magnet) prepared in a superposition state of various values of its classical parameter (here, two gradient directions)?}
    \label{fig:sketch}
\end{figure}

\section{The schemes}\label{unitary}

\subsection{Coherent scheme, exploiting the Zeno effect}

We start from a Hamiltonian for the system $H_S(\chi)$ that depends on a parameter $\chi$ describing the device, usually taken as classical. In the list of examples above, $\chi$ is the position $X$ of a scatterer, the transmittivity $t$ of a beam-splitter, and the direction $\hat{n}$ of the gradient of the magnetic field. Moving to a quantum description of the device's parameter, we need the evolution generated by the joint interaction $H_{SD}$. This should be derived from first principles; however, we can get a good general guess by remembering that a device in its classical state is one that does not get entangled with the system (the beam-splitter, just to take that one, does not record the momentum kick from the particle). This idea hints to the form 
\ba\label{HSDgen}
H_{SD}=\int d\chi H_S(\chi)\otimes\ketbra{\chi}{\chi},
\ea with $\ket{\chi}$ the quantum state of the device associated with the classical value $\chi$ of the parameter. The general result \eqref{effH} is independent of this form. 

The system and device are initialised in the product state
\begin{equation}\label{initialstate}
    \rho_{SD}^{(0)} = \rho_{S}^{(0)}\otimes\ketbra{\phi}{\phi}.
\end{equation} The initial state of the system $\rho_S^{(0)}$ is arbitrary, while the initial state of the device $\ket{\phi}$ is assumed to be pure to keep the discussion of the Zeno freezing simple. We then let the system and device evolve under $H_{SD}$ for a short time $\Delta t$, after which we perform a measurement \textit{on the device} in a basis that contains $\ket{\phi}$ with the goal of achieving Zeno freezing of the device in that state. The sequence of short evolutions followed by Zeno measurement is repeated over time.

We assume that the periodicity $\Delta t$ of the Zeno measurement is much shorter than any relevant timescale in the dynamics induced by $H_{SD}$, so that each step is well approximated by \begin{equation}\label{1storder}
    \rho_{SD}(t+\Delta t) = \rho_{SD}(t) + i\Delta t\left[\rho_{SD}(t),H_{SD}\right],
\end{equation}
where we have set $\hbar = 1$. We further assume that the Zeno measurement happens on even shorter time than $\Delta t$, even when we send this to zero.

Having set the assumptions, let us solve the dynamics. Under the first step of free dynamics \eqref{1storder}, the initial state \eqref{initialstate} evolves into \begin{equation}\label{stateDt}
    \rho_{SD}(\Delta t) = \rho_{SD}^{(0)} + i\Delta t \left[\rho_{S}^{(0)}\otimes\ketbra{\phi}{\phi},H_{SD}\right].
\end{equation}
For the Zeno effect, we consider a two-outcome projective measurement on the device, with projectors $\ketbra{\phi}{\phi}$ and $\mathbb{1} - \ketbra{\phi}{\phi}$. After such an instantaneous measurement on the evolved state \eqref{stateDt}, the new state $\rho_{SD}^{(1)}$ is given by
\ba
    \frac{\rho_{SD}^{(1)}-\rho_{SD}^{(0)}}{i\Delta t} & =
    & \left(\rho_{S}^{(0)}\otimes\ketbra{\phi}{\phi}\right)H_{SD}\left(\mathbb{1}\otimes\ketbra{\phi}{\phi}\right)\nonumber\\&& - \left(\mathbb{1}\otimes\ketbra{\phi}{\phi}\right)H_{SD}\left(\rho_{S}^{(0)}\otimes\ketbra{\phi}{\phi}\right)\nonumber\\
    &=&\frac{\rho_{S}^{(1)}-\rho_{S}^{(0)}}{i\Delta t}\otimes\ketbra{\phi}{\phi},
\ea
with
\begin{equation}
    \frac{\rho_{S}^{(1)}-\rho_{S}^{(0)}}{i\Delta t} =\left[\rho_{S}^{(0)},\bra{\phi}H_{SD}\ket{\phi}\right].
\end{equation}
We observe that the probability of obtaining outcome $\mathbb{1} - \ketbra{\phi}{\phi}$ vanishes at first order in $\Delta t$; in this stroboscopic evolution, the device is back in its initial state $\ket{\phi}$ and the system is prepared in the state $\rho_{S}^{(1)}$, which is the initial state for the next iteration. By taking the limit $\Delta t\rightarrow 0$ while assuming that the Zeno measurement is even shorter (instantaneous), the stroboscopic evolution becomes a unitary evolution generated by the effective Hamiltonian
\begin{equation}\label{effH}
    H_{S,\text{eff}}(\phi) = \bra{\phi}H_{SD}\ket{\phi}.
\end{equation}

\subsection{An incoherent scheme}

The effective Hamiltonian we have just derived does not depend on the coherences between eigenstates of $H_{SD}$. In particular, if $H_{SD}$ is of the form \eqref{HSDgen}, we have found
\ba
H_{S,\text{eff}}(\phi)&=&\int d\chi\,p_\phi(\chi)\,H_S(\chi)\,,
\ea where $p_\phi(\chi)=|\braket{\phi}{\chi}|^2$.
The fact that $H_{S,\text{eff}}(\phi)$
depends only on the weights of $\ket{\phi}$ on the classical states, and not on the coherences between these states, suggests that the same evolution could be generated by preparing the device in an \textit{incoherent mixture} of classical states, rather than a superposition. Indeed, we are going to describe such a scheme, as well as the two main differences between the coherent and incoherent case.

Notice first that one cannot freeze a mixed state with the Zeno effect, so the scheme cannot be identical. Rather, we consider the following: at every step $\Delta t$, a classical state $\ket{\chi}$ is drawn with probability $p_\phi(\chi)$, where for simplicity of the narrative we supposed that there are finitely many values of $\chi$. The device is then instantaneously re-prepared in that state. Each individual system will perceive a different channel $U(\chi^{N})...U(\chi^{1})$, depending on the sequence $\vec{\chi}=\{\chi^{1},...,\chi^N\}$ that was drawn. Over many such realisations, if the information about the various $\vec{\chi}$ is discarded (or not available), the averaged channel for the system will be of the form $\bar{U}_S^{N}$ with
\ba\bar{U}_S=\sum_{j}p_{\phi}(\chi_j)U(\chi_j)&\approx&\mathbb{1}-i\Delta t\, H_{S,\text{eff}}(\phi)\,.
\ea
Thus, the same effective evolution is achievable by simply randomizing the classical parameters. A few differences are nonetheless worth highlighting. First, as we stated, in this incoherent model the unitary channel generated by $H_{S,\text{eff}}(\phi)$ is only an ensemble average over several rounds, each characterised by a different configuration $\vec{\chi}$; whereas in the coherent scheme, the same channel is obtained in each round. Second, the randomization of classical parameters consists in instantaneously flipping the device between various classical states every $\Delta t$. This could be very disruptive, especially if the classical states are very different among themselves. By contrast, the Zeno freezing, although currently beyond experimental feasibility, would be in itself minimally disruptive.

\section{Case Studies}

In this section, we discuss cases in which $H_{S,\text{eff}}$ is just $H_S(\chi_{\text{eff}})$, and cases in which we expect a qualitatively different Hamiltonian.

It is clear that $H_{S,\text{eff}}=H_S(\chi_{\text{eff}})$ if $H_S(\chi)=\chi h$, or more generally $H_S(\chi)=f(\chi) h$ for any sufficiently continuous $f$. This would be the case for a beam-splitter in superposition of different transmittivity values. For another example of this type, we consider the interaction between Rydberg atoms and cavity fields as reported in experiments like \cite{PhysRevLett.77.4887,Harochebook}. There, the fully quantum description is $H= g \sigma_z\otimes N$ where $\sigma_z=\ketbra{e}{e}-\ketbra{g}{g}$ is the atomic operator, $N$ is the number of photons in the cavity, and $g$ is the coupling energy. But one could look at one of the systems as a device influencing the other. In either case, there is no interest in setting up one of our schemes, because exactly the same effective evolution of the system could be obtained by tuning the coupling constant \footnote{Indeed, the basic physics of those experiments has sometimes been presented semi-classically as the atom introducing an ``effective index of refraction'' that influences the evolution of the cavity field (although in none of those experiments there was a Zeno freezing of the atom's state).}.

The \textit{Stern-Gerlach (SG) interaction}, however, presents a more interesting case study. A toy model of SG has to capture the fact that the system gets a momentum kick in the $\pm\hat{n}$ direction if the spin points in that same direction:
\ba
    H_{S}\left(\hat{n}\right)&=&g\sum_{s=\pm 1} s\left(\hat{n}\cdot\vec{p}\right)\otimes\frac{\mathbb{1}+s\left(\hat{n}\cdot\vec{\sigma}\right)}{2}\nonumber\\ &=& g \left(\hat{n}\cdot\vec{p}\right)\otimes\left(\hat{n}\cdot\vec{\sigma}\right),
\ea where again $g$ denotes a coupling constant. In this case, we can consider $H_{SD} = g\int d\hat{n}H_{S}\left(\hat{n}\right)\otimes\ketbra{\hat{n}}{\hat{n}}$, where the integration is over all $\hat{n}$ in the plane orthogonal to the propagation axis of the atom. If the magnet is initialised in the superposition state $\ket{\phi}_{D} = \alpha\ket{\hat{n}_{1}} + \beta\ket{\hat{n}_{2}}$, we obtain $H_{S,\text{eff}} = \abs{\alpha}^{2}H_{S}\left(\hat{n}_{1}\right) + \abs{\beta}^{2}H_{S}\left(\hat{n}_{2}\right)$. In particular, if $\hat{n}_1\perp\hat{n}_2$ and $|\alpha|=|\beta|=\frac{1}{\sqrt{2}}$, the effective Hamiltonian is
\ba\label{HSG}
H_{S,\text{eff}} &=&\frac{g}{2}\left(p_x\otimes\sigma_x+p_y\otimes\sigma_y\right)\,,
\ea which is invariant by rotation in the $x-y$ plane. This is very different from any of the $H_S(\hat{n})$ because all trace of directionality is lost. 

When complemented by a screen in the far field, a SG evolution constitutes a \textit{SG measurement} of the spin direction. The interaction \eqref{HSG} is invariant by rotation and can be interpreted as the simultaneous pointer measurement of $\sigma_x$ and $\sigma_y$ \cite{doi:10.1002/andp.201700388}. The magnet may be deflect the particle in all possible directions, and preferentially along the Bloch vector $\vec{m}$ of the system (assumed to lie in the $x-y$ plane). This hints to the possibility of metrological advantage. Unfortunately, this is not the case for the SG spin measurement. To see it, notice that any transverse momentum kick, however small, will be detectable in the far field. So we can equally implement the incoherent scheme for one single step $\Delta t$: drawing $\hat{n}$ with uniform prior distribution in each round would lead to a SG measurement with interaction \eqref{HSG}.

Another situation, in which $H_{S,\text{eff}}$ may prove very different from any of the $H_S(\chi_{\text{eff}})$, is that of putting a \textit{scatterer} in superposition of different positions. For instance, if $V(x-x_0)$ is a step (which has only diffusive states), by a suitable arrangement one could create $V_{\text{eff}}\propto V(-x-a)+V(x-a)$, which is a well and therefore has bound states. This scattering example can also illustrate the complexity, mentioned above, of implementing the incoherent scheme. Every $\Delta t$, the scatterer may have to be instantaneously moved from one position to another. The Zeno freezing, in spite of creating coherences that would not be visible, would nonetheless look like a more convenient scheme if one were to find concrete ways to realise it.

\section{Conclusions}
We have answered the question of how a probe system would perceive a device prepared and kept in a quantum superposition state of the values of its classical parameters. Under ideal conditions, we have found that the system evolves unitarily according to the effective Hamiltonian \eqref{effH}. We have described an incoherent model that achieves the same evolution on an ensemble, at the price of rapidly flipping between classical states of the device during the evolution. Finally, we have discussed examples in which the effective Hamiltonian does not resemble that associated with a classical state of the device.

\section*{Acknowledgments}

We thank Alessandro Bisio, Paolo Perinotti and Aephraim Steinberg for useful inputs.

This research is supported by the National Research Foundation and the Ministry of Education, Singapore, under the Research Centres of Excellence programme.

\bibliographystyle{unsrt}
\bibliography{References}

\begin{thebibliography}{10}

\bibitem{schrodinger1935present}
Erwin Schr{\"o}dinger.
\newblock The present status of quantum mechanics.
\newblock {\em Die Naturwissenschaften}, 23(48):1--26, 1935.

\bibitem{Leggett_2002}
A~J Leggett.
\newblock Testing the limits of quantum mechanics: motivation, state of play,
  prospects.
\newblock {\em Journal of Physics: Condensed Matter}, 14(15):R415--R451, apr
  2002.

\bibitem{Harochebook}
Serge Haroche and Jean-Michel Raimond.
\newblock {\em Exploring the Quantum: Atoms, Cavities and Photons}.
\newblock Oxford University Press, Oxford, 2006.

\bibitem{RevModPhys.85.1103}
David~J. Wineland.
\newblock Nobel lecture: Superposition, entanglement, and raising
  schr\"odinger's cat.
\newblock {\em Rev. Mod. Phys.}, 85:1103--1114, Jul 2013.

\bibitem{arndt2014testing}
Markus Arndt and Klaus Hornberger.
\newblock Testing the limits of quantum mechanical superpositions.
\newblock {\em Nature Physics}, 10(4):271, 2014.

\bibitem{RevModPhys.90.025004}
Florian Fr\"owis, Pavel Sekatski, Wolfgang D\"ur, Nicolas Gisin, and Nicolas
  Sangouard.
\newblock Macroscopic quantum states: Measures, fragility, and implementations.
\newblock {\em Rev. Mod. Phys.}, 90:025004, May 2018.

\bibitem{shayeghi2019matterwave}
Armin Shayeghi, Philipp Rieser, Georg Richter, Ugur Sezer, Jonas~H. Rodewald,
  Philipp Geyer, Todd.~J. Martinez, and Markus Arndt.
\newblock Matter-wave interference of a native polypeptide, 2019.

\bibitem{ockeloen2018stabilized}
CF~Ockeloen-Korppi, E~Damsk{\"a}gg, J-M Pirkkalainen, M~Asjad, AA~Clerk,
  F~Massel, MJ~Woolley, and MA~Sillanp{\"a}{\"a}.
\newblock Stabilized entanglement of massive mechanical oscillators.
\newblock {\em Nature}, 556(7702):478--482, 2018.

\bibitem{riedinger2018remote}
Ralf Riedinger, Andreas Wallucks, Igor Marinkovi{\'c}, Clemens L{\"o}schnauer,
  Markus Aspelmeyer, Sungkun Hong, and Simon Gr{\"o}blacher.
\newblock Remote quantum entanglement between two micromechanical oscillators.
\newblock {\em Nature}, 556(7702):473--477, 2018.

\bibitem{Deliceaba3993}
Uro{\v s} Deli{\'c}, Manuel Reisenbauer, Kahan Dare, David Grass, Vladan
  Vuleti{\'c}, Nikolai Kiesel, and Markus Aspelmeyer.
\newblock Cooling of a levitated nanoparticle to the motional quantum ground
  state.
\newblock {\em Science}, 2020.

\bibitem{Schomerus_2002}
H~Schomerus, Y~Noat, J~Dalibard, and C.~W.~J Beenakker.
\newblock Multiple-path interferometer with a single quantum obstacle.
\newblock {\em Europhysics Letters ({EPL})}, 57(5):651--657, mar 2002.

\bibitem{PhysRevLett.96.173601}
Daniel Rohrlich, Yakov Neiman, Yonathan Japha, and Ron Folman.
\newblock Interference swapping in scattering from a nonlocal quantum target.
\newblock {\em Phys. Rev. Lett.}, 96:173601, May 2006.

\bibitem{PhysRevLett.115.260403}
Shi-Biao Zheng, You-Peng Zhong, Kai Xu, Qi-Jue Wang, H.~Wang, Li-Tuo Shen,
  Chui-Ping Yang, John~M. Martinis, A.~N. Cleland, and Si-Yuan Han.
\newblock Quantum delayed-choice experiment with a beam splitter in a quantum
  superposition.
\newblock {\em Phys. Rev. Lett.}, 115:260403, Dec 2015.

\bibitem{Angelo_2011}
Renato~M Angelo, Nicolas Brunner, Sandu Popescu, Anthony~J Short, and Paul
  Skrzypczyk.
\newblock Physics within a quantum reference frame.
\newblock {\em Journal of Physics A: Mathematical and Theoretical},
  44(14):145304, mar 2011.

\bibitem{giacomini2019quantum}
Flaminia Giacomini, Esteban Castro-Ruiz, and {\v{C}}aslav Brukner.
\newblock Quantum mechanics and the covariance of physical laws in quantum
  reference frames.
\newblock {\em Nature communications}, 10(1):1--13, 2019.

\bibitem{doi:10.1063/1.523304}
B.~Misra and E.~C.~G. Sudarshan.
\newblock The zeno’s paradox in quantum theory.
\newblock {\em Journal of Mathematical Physics}, 18(4):756--763, 1977.

\bibitem{chiribella2013quantum}
Giulio Chiribella, Giacomo~Mauro D’Ariano, Paolo Perinotti, and Benoit
  Valiron.
\newblock Quantum computations without definite causal structure.
\newblock {\em Physical Review A}, 88(2):022318, 2013.

\bibitem{brukner2014quantum}
{\v{C}}aslav Brukner.
\newblock Quantum causality.
\newblock {\em Nature Physics}, 10(4):259--263, 2014.

\bibitem{PhysRevLett.77.4887}
M.~Brune, E.~Hagley, J.~Dreyer, X.~Ma\^{\i}tre, A.~Maali, C.~Wunderlich, J.~M.
  Raimond, and S.~Haroche.
\newblock Observing the progressive decoherence of the ``meter'' in a quantum
  measurement.
\newblock {\em Phys. Rev. Lett.}, 77:4887--4890, Dec 1996.

\bibitem{Note1}
Indeed, the basic physics of those experiments has sometimes been presented
  semi-classically as the atom introducing an ``effective index of refraction''
  that influences the evolution of the cavity field (although in none of those
  experiments there was a Zeno freezing of the atom's state).

\bibitem{doi:10.1002/andp.201700388}
Nicolas Gisin and Emmanuel Zambrini~Cruzeiro.
\newblock Quantum measurements, energy conservation and quantum clocks.
\newblock {\em Annalen der Physik}, 530(6):1700388, 2018.

\end{thebibliography}
\end{document}